\documentclass{article}
\usepackage{graphicx,color}

\newcommand{\be}{\begin{equation}}
\newcommand{\ee}{\end{equation}}
\newcommand{\bdm}{\begin{displaymath}}
\newcommand{\edm}{\end{displaymath}}
\newcommand{\bea}{\begin{eqnarray}}
\newcommand{\eea}{\end{eqnarray}}

\newcommand{\bc}{\begin{center}}
\newcommand{\ec}{\end{center}}

\renewcommand{\hbar}{\rule[0.52em]{0.4em}{0.06em}\hspace{-0.45em}h}

\newcommand{\gs}{g_{\mbox{\tiny s}}}
\newcommand{\gw}{g_{\mbox{\tiny W}}}
\newcommand{\rw}{r_{\mbox{\tiny W}}}

\newcommand{\mw}{m_{\mbox{\tiny W}}}
\newcommand{\mz}{m_{\mbox{\tiny Z}}}

\begin{document}

\begin{center}
{\LARGE \bf Insights and puzzles in particle physics}\footnote{Contribution to {\it Fifty Years of Quarks}, H. Fritzsch and M. Gell-Mann, eds.\\ \rule{0.5cm}{0cm}(World Scientific, Singapore).}

\vspace{0.8cm}
H.~Leutwyler\\ Albert Einstein Center for Fundamental Physics\\Institute for Theoretical Physics, University of Bern\\
Sidlerstr.~5, CH-3012 Bern, Switzerland
\end{center}
 
 \begin{abstract}
I briefly review the conceptual developments that led to the Standard Model and discuss some of its remarkable qualitative features. On the way, I draw attention to several puzzling aspects that are beyond the reach of our present understanding of the basic laws of physics.
\end{abstract}
\tableofcontents
\section{Prehistory}\label{prehistory}
Bohr's model of the hydrogen atom (1913) gave birth to quantum theory and eventually led to a very thorough understanding of the structure of atoms, molecules, solids, \ldots\, In that framework, the electrons and the nuclei represent the constituents of matter. Their properties are controlled by mass and charge -- size and structure of the nuclei, magnetic moments etc. manifest themselves only in fine details of the picture.  

The first hint at the existence of particles other than electrons and nuclei occurred in $\beta$-decay, where the distribution of the decay products was puzzling because it violated energy conservation. Pauli solved the puzzle (1930): the observed spectrum can be explained if a yet unknown particle is emitted together with the electron. It must be neutral and escape detection. 
The experimental proof of existence for this particle became possible only much later. As Pauli wrote in his answer to the announcement of the discovery  (Reines and Cowan 1956):
 {\it Everything comes to him who knows how to wait.}

In today's terminology, Pauli predicted the electron neutrino, $\nu_e$. In fact, we now know that both the electron and the neutrino have relatives, leptons, which come in three families: 
\begin{center}\{$e$(1897), $\nu_e$(1956)\}\hspace{0.4cm}\{$\mu$(1936), $\nu_\mu$(1962)\}\hspace{0.4cm}\{$\tau$(1975), $\nu_\tau$(2000)\}. \end{center}
When the $\mu$ was discovered, Rabi asked: {\it Who ordered that ?} The existence of yet another charged lepton, the $\tau$, did not shed any light on this puzzle: why are there three families of leptons ?  

The discovery of the neutron (Chadwick 1932) simplified the picture considerably, as it reduced the number of constituents from over 90 to only 3: electron, proton, neutron. At the same time, it gave rise to a new puzzle: what forces the protons and neutrons to form nuclei ?  

Yukawa \cite{Yukawa} and St\"uckelberg  \cite{Stueckelberg} realized that the Coulomb force $V\sim \frac{1}{r}$ is of long range because it is due to the exchange of massless particles, photons. They noticed that the exchange of a particle of mass $m$ would instead give rise to a potential of finite range, $V\sim \frac{1}{r}e^{-r/r_0}$ and that the range is determined by the mass of the exchanged particle, $r_0=\hbar/mc$. While St\"uckelberg considered massive particles of spin 1, Yukawa investigated the exchange of massive particles with spin 0. From the fact that the range of the nuclear force is of the order of a few fermi, he predicted the existence of a spinless particle, which strongly interacts with protons and neutrons and has a mass of the order of 100 MeV$/c^2$. As this is intermediate between the masses of electron and proton or neutron, he coined the term {\it meson} for this particle. 

More than ten years later, the object was indeed found (Powell 1947), at a mass of about 140 MeV/$c^2$. It is now referred to as the $\pi$-meson or {\it pion}. Around the same time, many other strongly interacting particles started showing up: $K$-mesons, hyperons, excited states of the nucleon, \ldots

Gradually, the understanding of the nuclear forces developed into a very successful framework based on nonrelativistic quantum mechanics, where the interaction among the nucleons is described by means of a refined version of the Yukawa potential. The structure of the nuclei, nuclear reactions, the processes responsible for the energy production in the sun, $\alpha$-decay, etc. can all be understood on this basis.
 These phenomena concern interactions among nucleons with small relative velocities.   Experimentally, it had become possible to explore relativistic collisions. A description in terms of nonrelativistic potentials cannot cover these.  

\section{Situation at the beginning of the 1960ies}\label{1960}

General principles like Lorentz invariance, causality and unitarity had given deep insights: analyticity, dispersion relations, CPT theorem, relation between spin and statistics, for instance. 
Motivated by the successful predictions of Quantum Electrodynamics (QED), many
attempts at formulating a theory of the strong interaction based on elementary fields for baryons and mesons were undertaken,
but absolutely nothing worked even halfway. There was considerable progress in renormalization theory, but faith in quantum field theory was in decline, even concerning QED.  I illustrate the situation at the beginning of the 1960ies with the following quotations from Landau's assessment \cite{Landau}:

$\bullet$   {\it We are driven to the conclusion that the Hamiltonian method for strong interaction is dead
and must be buried, although of course with deserved honor.}

$\bullet$  {\it By now the nullification of the theory is tacitly
accepted even by theoretical physicists who profess to dispute it. This is evident from
the almost complete disappearance of papers on meson theory and particularly from
Dyson's assertion that the correct theory will not be found in the next hundred years.}
 
\noindent The basis of this pessimistic conclusion is clearly spelled out:

$\bullet$  {\it \ldots the effective interaction always diminishes with decreasing energy, so that the physical interaction at finite energies is always less than the interaction at energies of the order of the cut-off limit which is given by the bare coupling constant appearing in the Hamiltonian.}

\noindent In other words, all of the models of quantum field theory explored by that time had a positive $\beta$-function at weak coupling. Lagrangians involve products of the fields and their derivatives at the same space-time point. If the interaction grows beyond bounds when the distance shrinks, the Lagrangian is a questionable notion (for the asymptotically free theories to be discussed below, the $\beta$-function is negative at weak coupling, so that the problem does not arise, but these were discovered only later).

Many people doubted that the strong interaction could at all be described by means of a local quantum field theory. As a way out, it was suggested to give up quantum field theory and only rely on S-matrix theory -- heated debates about this suggestion took place \cite{Pietschmann}.  In short, fifty years ago, a theory of the strong interaction was not in sight. What was available was a collection of beliefs, prejudices and assumptions which where partly contradicting one another. As we now know, quite a few of these were wrong. The remaining ones are still with us \ldots

 \section{Quarks}\label{quarks}
 The Bohr Model (1913) played a key role in unraveling the structure of the atoms. The discovery of the Quark Model (1964) represents the analogous step in the developments which led to a solution of the puzzle posed by the strong interaction. Indeed, our understanding of the laws of nature made remarkable progress in the eight years between that discovery and the formulation of Quantum Chromodynamics (1972), the keystone of the Standard Model. 

Gell-Mann \cite{Gell-Mann SU3} and Ne'eman \cite{Ne'eman} independently pointed out that the many baryons and mesons observed by the beginning of the 1960ies can be grouped into multiplets which form representations of an approximate symmetry. The proposal amounts to an extension of isospin symmetry, which is characterized by the group SU(2), to the larger group SU(3). As indicated by the name {\it Eightfold Way}, the Lie algebra of SU(3) contains eight independent elements, which play a role analogous to the three components of isospin. 

On this basis, Gell-Mann predicted the occurrence of a baryon $\Omega^-$, needed to complete the decuplet representation of the Eightfold Way. In contrast to the other members of the multiplet, which represent rapidly decaying resonances, this particle is long-lived and thus leaves a trace in bubble chamber pictures. In 1964, the $\Omega^-$ was indeed found at the predicted mass, in a Brookhaven experiment. 
 
Zweig and Gell-Mann \cite{Zweig,Gell-Mann Quarks} then independently discovered that the multiplet pattern can qualitatively be understood if the strongly interacting particles are bound states formed with constituents of spin $\frac{1}{2}$, which transform according to the fundamental representation of SU(3): $u,d,s$. Zweig thought of them as real particles and called them `aces'. In view of the absence of any experimental evidence for such constituents, Gell-Mann was more reluctant.\footnote{{\it Such particles [quarks] presumably are not real but we may
use them in our field theory anyway \ldots} \cite{Gell-Mann Physics}}
He introduced the name `quark', borrowed from James Joyce\footnote{According to a story told to me by Lochlainn O'Raifeartaigh, James Joyce once visited an agricultural exhibit in Germany. There he saw the advertisement "Drei Mark f\"ur Musterquark" ({\it Mark} was the currency used in Germany at the time, {\it Quark} is the German word for cottage cheese, {\it Muster} stands for `exemplary', `model' or `sample'). Joyce is said to have been fond of playing around with words and may have come up with the famous passage in Finnegans Wake in this way: {\it Three quarks for Muster Mark! $|$ Sure he has not got much of a bark $|$
And sure any he has it's all beside the mark.}} and suggested to treat the quarks like the veal in one of the recipes practiced in the Royal french cusine (the pheasant was baked between two slices of veal, which were then left for the less royal members of the court). 

The Quark Model was difficult to reconcile with the spin-statistics theorem which implies that particles of spin $\frac{1}{2}$ must obey Fermi statistics. Greenberg proposed that the quarks obey neither Fermi-statistics nor Bose-statistics, but represent ``para-fermions of order 3'' \cite{Greenberg}. The proposal amounts to the introduction of a new internal quantum number. Indeed, in 1965, Bogoliubov, Struminsky and Tavkhelidze \cite{BST}, Han and Nambu \cite{HN} and Miyamoto \cite{Miyamoto} independently pointed out that some of the problems encountered in the quark model disappear if the $u$, $d$ and $s$ quarks occur in 3 different states. Gell-Mann coined the term ``colour'' for the new quantum number. 
 
 Today, altogether six quark {\it flavours} are needed to account for all of the observed mesons and baryons:  
\begin{center}\{$u$(1964), $d$(1964)\}\hspace{0.4cm}\{$c$(1974), $s$(1964)\}\hspace{0.4cm}\{$t$(1995), $b$(1977)\}. \end{center}
They also come in three families, but in contrast to the leptons, each one of the quarks comes in three versions, distinguished by the colour quantum number.

In 1968 Bjorken pointed out that if the nucleons contain point-like constituents, then the $ep$ cross section should obey scaling laws in the deep inelastic region \cite{Bjorken}. Indeed, the scattering experiments carried out by the MIT-SLAC collaboration in 1968/69 did show experimental evidence for such constituents \cite{Friedman Kendall Taylor}. Feynman called these {\it partons}, leaving it open whether they were the quarks or something else.

The operator product expansion turned out to be a very useful tool for the short distance analysis of the theory -- the title of the paper where it was introduced \cite{Wilson}, ``Non-Lagrangian models of current algebra'', reflects the general skepticism towards Lagrangian quantum field theory discussed in section \ref{1960}.

\section{Gauge fields}
According to the Standard Model, all interactions except gravity are mediated by the same type of fields: gauge fields. The electromagnetic field is the prototype. The final form of the laws obeyed by this field was formulated by Maxwell, around 1860 -- his formulation survived relativity and quantum theory, unharmed. While for the electrons, the particle aspect showed up first, the wave aspect of the e.m.\,field was thoroughly explored before the corresponding quanta, the photons, were discovered. 

Fock pointed out that the Schr\"odinger equation for electrons in an electromagnetic field is invariant under a group of local transformations \cite{Fock}. Weyl termed these  {\it gauge transformations}.  In fact, gauge invariance and renormalizability fully determine the form of the e.m.\,interaction. The core of Quantum Electrodynamics -- photons and electrons -- illustrates this statement. 
Gauge invariance and renormalizability allow only two free parameters in the Lagrangian of this system: $e,m_e$. 
Moreover, only one of these is dimensionless: $e^2/4\pi=1/137.035\, 999\, 074\, (44)$.  This shows that gauge invariance is the crucial property of the e.m.\,interaction: together with renormalizability, it fully determines the properties of the e.m.~interaction, except for this number, which so far still remains unexplained.

The symmetry group that characterizes the electromagnetic field is the group U(1), but gauge invariance can be generalized to larger groups, such as SU(2) or SU(3) \cite{Yang and Mills,Shaw}. Gauge invariance then requires the occurrence of more than one gauge field: 3 in the case of SU(2), 8 in the case of SU(3), while a single gauge field is needed for U(1). Pauli had encountered this generalization of the concept of a gauge field earlier, when extending the Kaluza-Klein scenario to Riemann spaces of more than five dimensions. He did not consider this worth publishing, however, because he was convinced that the quanta of a gauge field are necessarily massless like the photon: gauge invariance protects these particles from picking up mass. In particular, inserting a mass term in the Lagrangian is not allowed, because such a term violates gauge invariance. Pauli concluded that the forces mediated by gauge fields are of long range and can therefore not possibly describe the strong or weak interactions -- these are of short range \cite{Straumann}.

Ten years later, Englert and Brout \cite{Englert and Brout}, Higgs \cite{Higgs},  and Guralnik, Hagen and Kibble \cite{Guralnik et al} showed that Pauli's objection is not valid in general: in the presence of scalar fields, gauge fields can pick up mass, so that forces mediated by gauge fields can be of short range. The work of Glashow \cite{Glashow}, Weinberg \cite{Weinberg} and Salam \cite{Salam} then demonstrated that non-abelian gauge fields are relevant for physics: the framework discovered by Englert et al.~does lead to a satisfactory theory of the weak interaction. 
\section{Quantum Chromodynamics}
One of the possibilities considered for the interaction that binds the quarks together was an abelian gauge field analogous to the e.m.~field, but this gave rise to problems, because the field would then interfere with the other degrees of freedom. Fritzsch and Gell-Mann pointed out that if the gluons carry colour, then the empirical observation that quarks appear to be confined might also apply to them: the spectrum of the theory might exclusively contain colour neutral states. 

Gell-Mann's talk at the High Energy Physics Conference in 1972 (Fermilab), had the title ``Current algebra: Quarks and what else?'' In particular, he discussed the proposal to describe the gluons in terms of a {\it non-abelian gauge field} coupled to colour, relying on work done with Fritzsch \cite{Gell-Mann Fermilab}.  As it was known already that the electromagnetic and weak interactions are mediated by gauge fields, the idea that colour might be a local symmetry as well does not appear as far fetched. In the proceedings, Fritzsch and Gell-Mann mention unpublished work in this direction by Wess. 

The main problem at the time was that for a gauge field theory to describe the hadrons and their interaction, it had to be fundamentally different from the quantum field theories encountered in nature so far. All of these, including the electroweak theory, have the spectrum indicated by the degrees of freedom occurring in the Lagrangian: photons, leptons, intermediate bosons, \ldots\,The Lagrangian of the strong interaction can be the one of a gauge field theory only if the spectrum of physical states of a quantum field theory can be qualitatively different from the spectrum of fields needed to formulate it: gluons and quarks in the Lagrangian, hadrons in the spectrum. In 1973, when the arguments in favour of QCD as a theory of the strong interaction were critically examined \cite{Fritzsch Gell-Mann Leutwyler}, the idea that the observed spectrum of hadrons can fully be understood on the basis of a theory built with quarks and gluons still looked rather questionable and was accordingly formulated in cautious terms. That this is not mere wishful thinking became clear only when the significance of the fact that the $\beta$-function of a non-abelian gauge field theory is negative at weak coupling was recognized. Interactions with this property are called {\it asymptotically free}. The following few comments refer to the history of this concept.\footnote{I thank Mikhail Vysotsky, Zurab Silagadze, Martin L\"uscher, J\"urg Gasser and Chris Korthals-Altes for information about this story.}  Further historical material concerning developments relevant for QCD is listed in reference \cite{history QCD}. 

Already in 1965, Vanyashin and Terentyev \cite{Vanyashin and Terentyev} found that the renormalization of the electric charge of a vector field is of opposite sign to the one of the electron (the numerical value of the coefficient was not correct). In the language of SU(2) gauge field theory, their result implies that the $\beta$-function is negative at one loop.  

The first correct calculation of the $\beta$-function of a non-abelian gauge field theory was carried out by Khriplovich, for the case of SU(2), relevant for the electroweak interaction \cite{Khriplovich}. He found that $\beta$ is negative and concluded that the interaction becomes weak at short distance. 

In his PhD thesis, 't Hooft performed the calculation of the $\beta$-function for an arbitrary gauge group, including the interaction with fermions and Higgs scalars \cite{'t Hooft}. He proved that the theory is renormalizable and confirmed that, unless there are too many fermions or scalars, the $\beta$-function is negative at weak coupling. 

This demonstrates that there are exceptions to Landau's rule, according to which {\it the effective interaction always diminishes with decreasing energy}: non-abelian gauge theories show the opposite behaviour, asymptotic freedom.   

Symanzik pointed out that for theories with a negative $\beta$-function at weak coupling, the behaviour of the Green functions at large momenta is controlled by perturbation theory and hence computable \cite{Symanzik}. The dimensions of the field operators are then the same as for free fields; the interaction only generates corrections that disappear at high energies, in inverse proportion to the logarithm of the momentum. He presented his results at a workshop in Marseille in 1972 \cite{Marseille} and in the discussion that followed his talk, t'Hooft pointed out that non-abelian gauge theories can have a negative $\beta$-function at weak coupling. 

Parisi discussed the consequences of a negative $\beta$-function for the structure functions of deep inelastic scattering and suggested that this might explain Bjorken scaling \cite{Parisi}. Gross and Wilczek \cite{Gross and Wilczek} and Politzer \cite{Politzer} then showed that -- if the strong interaction is mediated by a non-abelian gauge field -- the asymptotic behaviour of the structure functions is indeed computable. They predicted specific logarithmic modifications of the scaling laws. In the meantime, there is strong experimental evidence for these. 
 
\section{Standard Model}

In the Standard Model, the interactions among the constituents of matter are generated by  three distinct gauge fields:
\begin{center}
 \begin{tabular}{|lr|c|c|l|c|l|}\hline
interaction&&group&dim.&particles& source&coupling\\
\hline
electromagnetic&QED& U(1)&1&photon&charge&\hspace{1.2em}$e$ \\
weak&QFD&SU(2)&3&W$^+$\,W$^-$\,Z&flavour&\hspace{1.2em}$\gw$ \\
strong&QCD &SU(3)&8&gluons& colour&\hspace{1.2em}$\gs$ \\
\hline
\end{tabular}\end{center}
The symmetry which underlies the weak interaction manifests itself in the fact that the families come in doublets: $\{e,\nu_e\}$, $\{u,d\}$, \ldots\hspace{0.07cm}The symmetry group SU(2), which characterizes the corresponding gauge field theory, Quantum Fla\-vourdynamics (QFD), takes the electron into a mixture of an electron and a neutrino, much like an isospin rotation mixes the members of the isospin doublet formed with the proton and the neutron. More precisely, the gauge transformations of QFD only affect the left-handed components of the doublets and leave the right-handed components alone. We do not know why that is so, nor do we understand why there is no gauge symmetry connecting the different families. 

The symmetry group SU(3) acts on the colour of the quarks, mixing the three colour states. Since the leptons are left alone by the gauge group that underlies QCD, they do not participate in the strong interaction.  

The statement that all of the interactions except gravity are mediated by the same type of field sounds miraculous.  
Since a long time, it is known that the electromagnetic, weak and strong interactions have qualitatively very different properties.
How can that be if all of them are generated by the same type of field ? For this puzzle, the Standard Model does offer an explanation -- the following sections address this question.

\section{Why is QED different from QCD ?}
I first discuss the origin of the difference between the electromagnetic and strong interactions. It originates in the fact that the photons are electrically neutral while the gluons are coloured objects, transforming according to the octet representation of the colour group. The mathematical reason for the difference is that the gauge group of Quantum Electrodynamics, U(1), is an abelian group while the gauge group of Quantum Chromodynamics, SU(3), is non-abelian: for $x_1, x_2\in$ U(1) the product  $x_1\cdot x_2$ is identical with $x_2\cdot x_1$, but for  $x_1, x_2\in$ SU(3), the element $x_1\cdot x_2$ in general differs from $x_2\cdot x_1$.
 
The two plots below compare the surrounding of a lepton with the one of a quark. In either case, the interaction polarizes the vacuum. 
A positron generates a cloud of photons in its vicinity as well as a cloud of electrons and positrons. The electrons dominate, so that the charge density, which is shown in the plot on the left, receives a contribution that is of opposite sign to the bare charge inserted in the vacuum. Accordingly, the total charge of the positron is smaller than the bare charge, $e<e\hspace{0.1em}\rule[-0.3em]{0.05em}{0.8em}_{\,\mbox{\tiny\sffamily bare}}$: the vacuum shields the charge.

Likewise, a bare quark is surrounded by a cloud of gluons, quarks and antiquarks. The difference between QED and QCD manifests itself in the properties of these clouds: the photon cloud (the Coulomb field surrounding the bare positron) does not contribute to the charge density, because the photons are electrically neutral, but the gluons do show up in the density of colour, because they are coloured. The corresponding contribution to the colour density turns out to be of the same sign as the one of the bare quark at the center. The gluon cloud thus amplifies the colour rather than shielding it. 

\vspace{0.5cm}\hspace{-0.3cm}
\includegraphics[width=5cm]{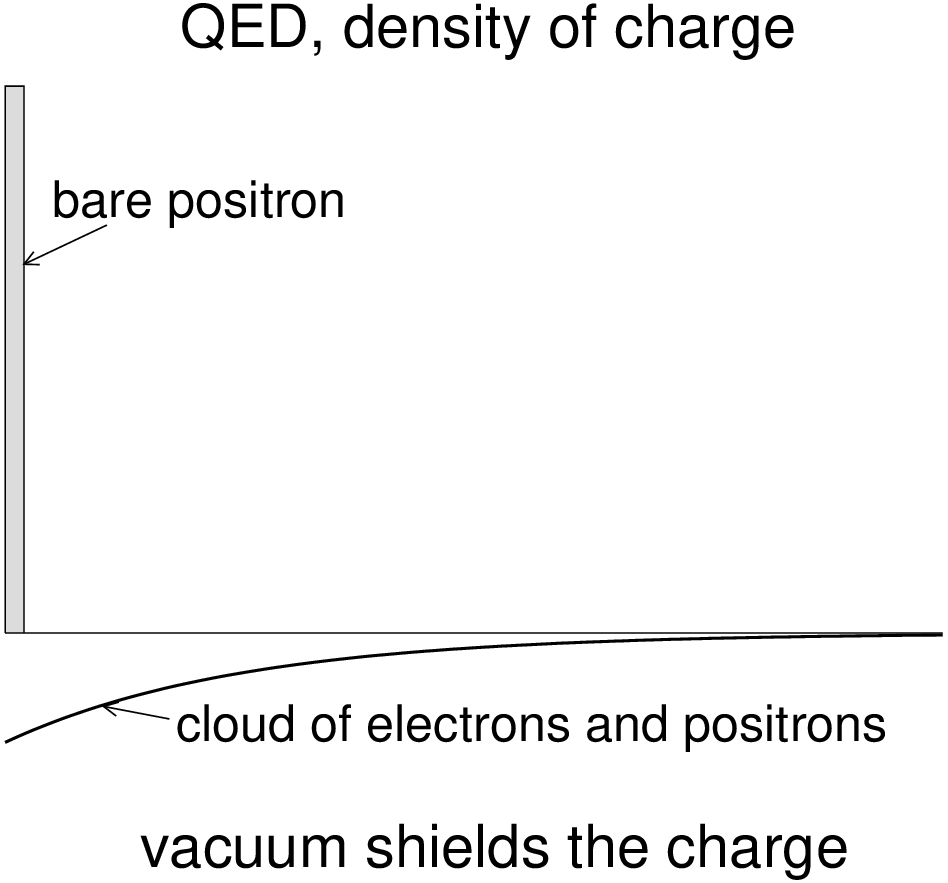}\hspace{1.5cm} 
\includegraphics[width=5cm]{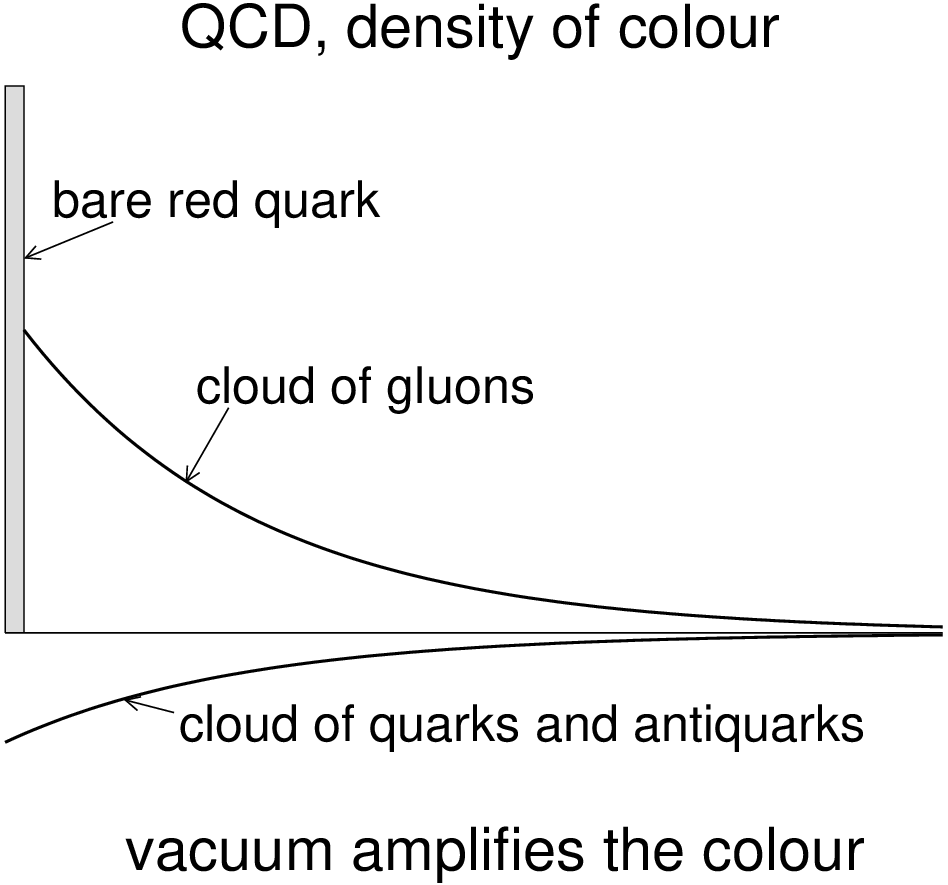}

\vspace{0.3cm}
Concerning the cloud of quarks and antiquarks, the situation is the same as in the plot on the left: their contribution to the colour density is of opposite sign to the one from the bare quark. Unless there are too many quark flavours, the contribution from the gluons is more important than the one from the quarks and antiquarks: the vacuum amplifies the colour. As it is the case with QED, vacuum polarization implies that the effective strength of the interaction depends on the scale, but while $e$ shrinks with increasing size of the region considered, $g_s$ grows:  $\gs>\gs\hspace{0.05em}
\rule[-0.4em]{0.05em}{0.8em}_{\,\mbox{\tiny\sffamily bare}}$. 
In other words, the qualitative difference between QED and QCD arises because the two interactions polarize the vacuum differently.
\section{Comparison with gravity }
A qualitatively very similar effect also occurs in gravity. The source of the gravitational field is the energy.  The gravitational field itself carries energy, comparable to the situation in Chromodynamics, where the colour is the source of the field and the gluons act as their own source, because they carry colour. In application to the motion of a planet like Mercury, the planet is attracted only by the energy contained within its orbit -- the forces generated by the gravitational field outside cancel out. This produces a tiny effect in the perihelion shift: if the orbit is calculated under the assumption that the attraction is produced by the total energy of the sun, the perihelion shift comes out larger than predicted by general relativity: 50" instead of 43" per century. 

The figure compares the orbit of Mercury with the charged pion, where two quarks are orbiting one another and where the effect is much stronger. Unfortunately, a quantitative comparison cannot be made, because gravity can be compared with Chromodynamics only at the classical level -- a quantum field theory of gravity is not available and a description of the structure of the pion within classical field theory is not meaningful.   

\begin{center} \includegraphics[width=6.5cm]{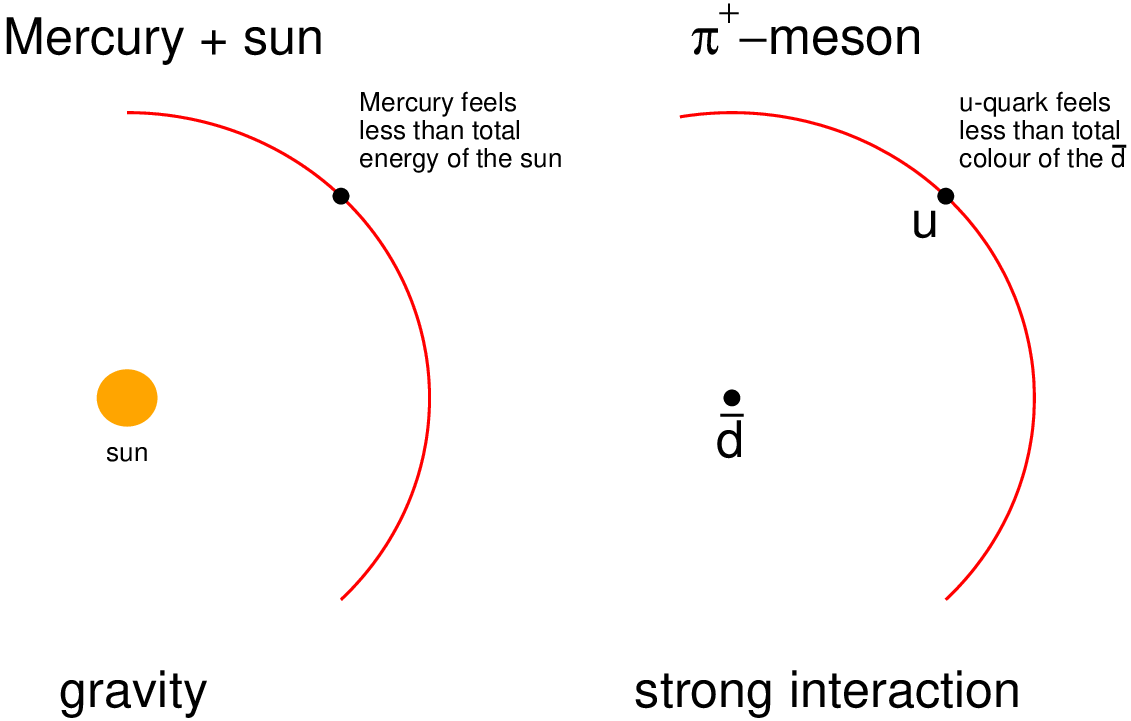}\end{center}

\section{Physics of vacuum polarization}
The fact that the vacuum reduces the electric field of a charged source but amplifies the gluonic field of a coloured source has dramatic consequences: although the Lagrangians of QED and QCD are very similar,
the properties of the electromagnetic and strong interactions are totally
different. 

The most important consequence of vacuum polarization is that QCD confines colour while QED does not confine charge. Qualitatively, this is easy to understand: the energy density of the e.m.\,field that surrounds a positron falls off with the distance -- the field energy contained in the region outside a sphere of finite radius is finite. Accordingly, only a finite amount of energy is needed to isolate a positron from the rest of the world: charged particles can live alone, the electric charge is not confined. For a quark, the situation is different, because the gluonic field that surrounds it does not fall off with the distance. In order to isolate a quark from the rest of the world, an infinite amount of energy would be needed. Only colourless states, hadrons, can live alone: mesons, baryons, nuclei.    

As Zweig pointed out already in 1964, the forces between the colourless objects can be compared with the van der Waals forces between atoms \cite{Zweig}. At long distance and disregarding the e.m.\,interaction, the force between two hadrons is dominated by the exchange of the lightest strongly interacting particle, the $\pi$-meson: the Yukawa formula is indeed valid in QCD with $r_0=\hbar/m_\pi c$, not for the force between two quarks, but for the force between two hadrons. It is of short range because the pion is not massless. The van der Waals force between two atoms, on the other hand, is of long range as it only drops off with a power of the distance: atoms can exchange photons and these are massless.  
 
An analytic proof that QCD does indeed confine colour is not available, but the results obtained by means of numerical simulations of QCD on a lattice represent overwhelming evidence that this is the case. The progress made with such simulations in recent years has made it possible to calculate the masses of the lowest lying bound states of QCD from first principles, in terms of the parameters occurring in the Lagrangian. The fact that the result agrees with the observed mass spectrum provides a very thorough test of the hypothesis that the quarks are bound together with a gauge field. 

\section{Why is QFD different from QCD ?}
The above discussion shows that the difference between QED and QCD arises from vacuum polarization, but this cannot explain why the weak and strong interactions are so different. The vacuum amplifies flavour for the same reason as it amplifies colour: both of these theories are asymptotically free. Why, then, is colour confined, flavour not ?

The difference arises because, in addition to the constituents and the gauge fields that mediate the interaction between them, the Standard Model contains a third category of fields: {\it scalar fields}, referred to as {\it Higgs fields}. These affect the behaviour of the weak interaction at long distance, but leave the strong and electromagnetic interactions alone. As shown by Glashow, Salam and Weinberg, an SU(2) doublet of Higgs fields is needed to describe the properties of the weak interaction: four real scalar fields are required. 

Lorentz invariance allows scalar fields to pick up a vacuum expectation value. In particle terminology: spinless particles can form a {\it condensate}. For the gauge fields which mediate the weak interaction, the condensate feels like a medium which affects the $W$ and $Z$ bosons that fly through it. 

If the frequency of a $W$-wave is less than a certain critical value $\omega_{\mbox{\tiny W}}$, which is determined by the strength of the interaction between the $W$ and the Higgs fields, it penetrates only a finite distance into the medium. For waves of low frequency, the penetration depth is given by $\rw=c/\omega_{\mbox{\tiny W}}$. If the frequency is higher than $\omega_{\mbox{\tiny W}}$, then the wave can propagate, but the group velocity is less than the velocity of light -- the condensate impedes the motion. In particle language: the $W$-particles pick up mass. The size of the mass is determined by the critical frequency, $\mw=\hbar \hspace{0.1em}\omega_{\mbox{\tiny W}}/c^2$. 

In the presence of the condensate, the interaction mediated by the $W$-bosons thus involves the exchange of massive particles. Accordingly, the $\frac{1}{r}$-potential characteristic of the exchange of massless particles  is modified:
\bdm
  \frac{\gw^2}{ 4\pi r}\hspace{0.2em}\Rightarrow \hspace{0.2em}
\frac{\gw^2}{4\pi r}\cdot  \mbox{\large$ e$}^{-\mbox{\large $ \frac{r}{\rw}$}}\hspace{1cm} \rw=\frac{\hbar}{\mw c}\,.\edm
The condensate thus gives the weak interaction a finite range, determined by the penetration depth.  At low energies, only the mean value of the potential counts, which is given by 
\bdm\int\! d^3 r\; \frac{\gw^2}{\rule{0em}{0.6em}4\pi r}\cdot\mbox{\large$ e$}^{-\mbox{\large $ \frac{r}{\rw}$}}= 
\gw^2\rw^2\edm
In the Fermi theory of 1934 \cite{Fermi}, the weak interaction is described by the contact potential $V_F=4\sqrt{2}\, G_F\delta^3(\vec{x})$, for which the mean value is given by $4\sqrt{2}\, G_F$. At low energies, the exchange of $W$-bosons thus agrees with the Fermi theory, provided $\gw^2\rw^2=4\sqrt{2}\, G_F$. 

Experimentally, the gauge particles which mediate the weak interaction were discovered at the CERN SPS (UA1 and UA2) in 1983. The mass of the $W$ turns out to be larger than the mass of an iron nucleus:
$\mw=  85.673\pm0.016$ $m_{\mbox{\tiny proton}}$. The $Z$, which is responsible for the weak interaction via neutral currents, is even heavier: $\mz=97.187\pm0.002$ $m_{\mbox{\scriptsize proton}}$. This explains why the range of the weak interaction is very short, of the order of $2\cdot10^{-18}\,$m, and why, at low energies, the strength of this interaction is so weak.

Massless gauge particles have only two polarization states, while massive particles of spin 1 have three independent states.
The Higgs fields provide the missing states: three of the four real Higgs scalars are eaten up by the gauge fields of the weak interaction. Only the fourth one survives.  Indeed a Higgs candidate was found in 2012, by the ATLAS and CMS teams working at the CERN LHC. As far as its properties revealed themselves by now, they are in accord with the theoretical expectations.

\section{Transparency of the vacuum}
The above discussion of the qualitative differences between the electromagnetic, weak and strong interactions concerns the behaviour at long distance. At short distance ($10^{-19}\,$m $\leftrightarrow$ 2 TeV), all of the forces occurring in the Standard Model obey the inverse square law. The interaction energy is of the form
\bdm \displaystyle V = \mbox{constant}\times
\frac{\hbar c}{r}\,,\edm
where the constant is a pure number. As the gauge symmetry consists of three distinct factors, $G=\mbox{U(1)}\times\mbox{SU(2)}\times\mbox{SU(3)}$, the strength of the interaction at short distances is characterized by three distinct numbers: 
\begin{center}\begin{tabular}{ccc}e.m.&\rule{1cm}{0cm}weak&\rule{1cm}{0cm}strong\\
 $\displaystyle\frac{e^2}{4\pi }$&\rule{1cm}{0cm}$\displaystyle\frac{\gw^2}{4\pi}$&\rule{1cm}{0cm}$\displaystyle\frac{\gs^2}{4\pi}$
\end{tabular}\end{center}
Vacuum polarization implies that each of the three constants depends on the scale. If the size of the region considered grows, then the e.m.\,coupling $e$ diminishes (the vacuum shields the charge), while the coupling constant $\gw$ of the weak interaction as well as the coupling constant $\gs$ of the strong interaction grows (the vacuum amplifies flavour as well as colour). 

At distances of the order of the penetration depth of the weak interaction, however, the presence of a condensate makes itself felt. Since the particles in the condensate are electrically neutral, the photons do not notice them -- for photons, the vacuum is a transparent medium and the mediation of the electromagnetic interaction is not impeded by it. The condensed particles do not have colour, either, so that the gluons do not notice their presence -- the vacuum is transparent also for gluons. The gauge field of the weak interaction, however, does react with the medium, because the condensed Higgs particles do have flavour: while the Higgs fields are invariant under the gauge groups U(1) and SU(3), they transform according to the doublet representation of SU(2). As discussed above, $W$ and $Z$-waves of low frequency cannot propagate through this medium: for such waves the vacuum is opaque. The amplification of the weak interaction due to the polarization of the vacuum only occurs at distances that are small compared to the penetration depth. At larger distances, the condensate takes over and shielding eventually wins -- although the weak interaction is asymptotically free, flavour is not confined.

This also implies that, at low energies, the weak interaction and the Higgs fields freeze out: the Standard Model reduces to QED+QCD.  In cold matter, only the degrees of freedom of the photon, the gluons, the electron and the $u$- and $d$-quarks manifest themselves directly. The remaining degrees of freedom only show up indirectly, through small, calculable corrections that are generated by the quantum fluctuations of the corresponding fields. 
 
\section{Masses of the leptons and quarks}
 The leptons and quarks also interact with the condensate. Gauge invariance does not allow the presence of corresponding mass terms in the Lagrangian, but they pick up mass though the same mechanism that equips the $W$ and $Z$-bosons with a mass. The size of the lepton and quark masses is determined by the strength of their interaction with the Higgs fields. Unfortunately, the symmetries of the Standard Model do not determine the strength of this interaction.
 
 Indeed, the pattern of lepton and quark masses is bizarre: the observed masses range
from $10^{-2}\,$eV/$c^2$ to $10^{+11}\,$eV/$c^2$. It so happens that the electron only interacts very weakly with the condensate and thus picks up only little mass. For this reason, the size of the atoms, which is determined by the Bohr radius, $\displaystyle a_{\mbox{\scriptsize Bohr}}=4\pi\hbar/e^2 m_ec$, is much larger than the size of the proton, which is of the order of the range of the strong interaction, $\hbar /m_\pi c$ -- but why is the interaction of the electrons with the Higgs particles so weak ?

\section{Beyond the Standard Model}
The Standard Model leaves many questions unanswered. In particular, it neglects the gravitational interaction. By now, quantum theory and gravity peacefully coexist for almost a century, but a theory that encompasses both of them and is consistent with what is known still remains to be found. This also sets an upper bound on the range of validity of the Standard Model: at distances of the order of the Planck length, $\ell_{\mbox{\tiny Planck}}=\sqrt{G\, \hbar/c^3}= 1.6 \cdot 10^{-35}\,$m, the quantum fluctuations of the gravitational field cannot be ignored. 

Neither the electromagnetic interaction nor the interaction among the Higgs fields is asymptotically free. Accordingly, the Standard Model is inherently incomplete, even apart from the fact that gravity can be accounted for only at the classical level. A cut-off is needed to give meaning to the part of the Lagrangian that accounts for those interactions. The Standard Model cannot be the full story, but represents an effective theory, verified up to  energies of the of order 1 TeV. It may be valid to significantly higher energies, but must fail before the Planck energy $E_{\mbox{\tiny Planck}}\sim 1.2\cdot 10^{\,28}\,$eV is reached. Quite a few levels of structure were uncovered above the present resolution. It does not look very plausible that there are no further layers all the way to $\ell_{\mbox{\tiny Planck}}$.

Astronomical observations show that the universe contains Dark Matter as well as Dark Energy. The latter may be accounted for with a cosmological constant, but we do not understand why this constant is so small. Anything that carries energy generates gravity. Why does gravity not take notice of the Higgs condensate ? The Standard Model does not have room for Dark Matter, either. Supersymmetric extensions of the Standard Model do contain candidates for Dark Matter, but so far, not a single member of the plethora of super-partners required by this symmetry showed up.  

We do not understand why the baryons dominate the visible matter in our vicinity. In fact, this is difficult to understand if the proton does not decay and until now, there is no upper bound on the proton lifetime. How could the observed excess of quarks over antiquarks have arisen if processes which violate baryon number do not occur ? In particular, CP violation is necessary for baryogenesis. The phenomenon is observed and can be accounted for in the Standard Model, but we do not understand it, either.

Why are there so many lepton and quark flavours ? What is the origin of the bizarre mass pattern of the leptons and quarks ?
The Standard Model becomes much more appealing if it is sent to the hairdresser, asking him or her to chop off all fields except the quarks and gluons. In the absence of scalar fields, a Higgs condensate cannot occur, so that the quarks are then massless: what survives the thorough cosmetic treatment is {\it QCD with massless quarks}. 

This is how theories should be: massless QCD does not contain a single dimensionless parameter to be adjusted to observation. 
 In principle, the values of all quantities of physical interest are predicted without the need to tune parameters (the numerical value of the mass of the proton in kilogram units cannot be calculated, because that number depends on what is meant by a kilogram, but the mass spectrum, the width of the resonances, the cross sections, the form factors, \ldots\hspace{0.07cm}can be calculated in a parameter free manner from the mass of the proton, at least in principle).  This theory does explain the occurrence of mesons and baryons and describes their properties, albeit only approximately -- for an accurate representation, the quark masses cannot be ignored.  
Compared to this beauty of a theory, the Standard Model leaves much to be desired \ldots

\end{document}